\documentclass [a4paper,12pt]{article}
\usepackage {graphicx,amsmath}
\usepackage{hyperref}
\usepackage{cleveref}  

\crefname{equation}{Eq.}{Eqs.}
\crefname{section}{Sec.}{Secs.}
\crefname{chapter}{Chapter}{Chapters}
\crefname{table}{Table}{Tables}
\crefname{figure}{Fig.}{Figs.}
\crefname{appsec}{Appendix}{Appendices}
\crefname{appchap}{Appendix}{Appendices}
\begin{document}
\vspace{-5mm}
\begin{center}
 {\Large\bf Two different physical interpretations of Schr\"{o}dinger equation } \\[2mm]
       Milo\v{s} V. Lokaj\'{\i}\v{c}ek, Vojt\v{e}ch Kundr\'{a}t and Ji\v{r}\'{\i} Proch\'{a}zka    
\footnote{e-mail: lokaj@fzu.cz, kundrat@fzu.cz, prochazka@fzu.cz} 
\\
{\it Institute of Physics of the AS CR, v.v.i., 18221 Prague 8, Czech Republic} \\
\end{center}

{\bf Abstract } \\
The assumptions added by Bohr and concerning the Hilbert space (formed by all solutions of Schr\"{o}dinger equation) changed fundamentally the original physical interpretation of these solutions proposed earlier by Schr\"{o}dinger. This new alternative was refused by Einstein on the basis of the EPR Gedankenexperiment, but accepted fully for microscopic reality by scientific community. Both the quantum alternatives were discussed, however, again later. Bell tried to find a possibility how to decide between them; he generalized Einstein's Gedankenexperiment assuming that also spins of two detected particles would be measured. He derived then some inequality for a special combination of four coincidence probabilities, and it was commonly assumed that his inequality held for the original Schr\"{o}dinger interpretation but not in Bohr's Copenhagen quantum mechanics; without any actual proof having been given. Corresponding experiments were proposed and finished in 1982. The violation of Bell's inequality was then interpreted as decisive victory of Bohr's theory. However, it will be shown that Bell's inequality has been interpreted mistakenly. It has been based always on some assumption that does not hold in any probabilistic theory (i.e., in the given spin experiment) but only in deterministic classical theory. There is not any argument for preferring the Copenhagen quantum mechanics and against Einstein's critical standpoint. Some other consequences will be mentioned, too.  
    \vspace{5mm}


\\

\section{\label{sec:introduction}Introduction}

The equation proposed by Schr\"{o}dinger in 1925 was accepted by physical community when it was shown that it might be interpreted in agreement with classical results \cite{schr}. However, this interpretation was fundamentally changed when some important assumptions were added by Bohr in 1927 (see \cite{bohr1}). It was then demonstrated by Einstein with the help of his Gedankenexperiment in 1935 that Bohr's Copenhagen quantum mechanics included the existence of immediate interaction between two very distant matter objects \cite{einst}, denoted as entanglement or non-locality at the present. Bohr \cite{bohr} refused Einstein's criticism, stating that the given phenomenon may exist in the microscopic world, which was fully accepted by world scientific community.  

This standpoint of the scientific community was influenced mainly by the conclusion of von Neumann \cite{vonne} having been presented by him in 1932 and refusing the existence of any hidden parameters in Schr\"{o}dinger equation; the problem having been discussed earlier. Consequently, the Copenhagen mechanics was accepted as the only theory of the microscopic world even if 
nobody was able to define a corresponding boundary between macroscopic and microscopic regions and even if G.~Herrmann \cite{gher} demonstrated in 1935 that the argument of von Neumann was to be denoted as circular reasoning. 

The situation changed partially in 1952 when Bohm \cite{bohm} showed that a hidden variable was contained in Schr\"{o}dinger equation even if its impact was not very clear. Two different interpretations of Schr\"{o}dinger equation started to be discussed for the region of the microscopic world: Copenhagen quantum mechanics (CQM) of Bohr and the other alternative denoted sometimes as bohmian (taken as corresponding to Einsteins's requirement). However, while the former alternative was defined exactly by Bohr's additional assumptions (corresponding Hilbert space being spanned always only on one set of Hamiltonian eigenfunctions and each vector of this limited Hilbert space representing a pure state) the exact definition of additional assumptions in the latter alternative has not been given.

Bell \cite{bell} tried to bring the decision between these two alternatives and derived his famous inequality in 1964. In deriving this inequality he generalized the purely classical Gedankenexperiment proposed originally by Einstein in which only coincidence detection of two particles formed by the decay of a particle in the rest and running in opposite directions was assumed to be established. Bell added also spin characteristics and assumed that the decaying particle had zero spin while both the decay particles possessed non-zero spins (oriented in opposite directions); he assumed further that also the coincidence probabilities of spin orientations of individual particles would be measured:
    \[ \left\|<--|^{\beta}---o---|^{\alpha}-->\right\|  \]
 He introduced then the following sum of four coincidence probabilities
\begin{equation}  
      B =  a_{\alpha_1}b_{\beta_1} + a_{\alpha_1}b_{\beta_2} + a_{\alpha_2}b_{\beta_1} 
                                      - a_{\alpha_2}b_{\beta_2}       \label{bellprob}  
\end{equation}
where $a_{\alpha_1}, a_{\alpha_2}, \; b_{\beta_1},b_{\beta_2}  \leq 1$ represented measured probabilities of particle transmissions through two polarizers (one polarizer and one detector on each side) at four different orientations (characterized by angles $\alpha_1$, $\alpha_2$, $\beta_1$ and $\beta_2$).  If the opposite spins of the decay particle pair were distributed randomly Bell derived for the given probability combination to hold always  
\begin{equation}  
       B\; \leq \;  2\,.  
\end{equation} 
It was then commonly assumed (without any proof) that this Bell inequality was  valid in the latter (bohmian) alternative, but not in the former one (CQM). 

The experiments making possible to measure corresponding spin orientations were proposed and performed in 1982; the coincidence probabilities of two photons emitted by excited atom in zero spin state and passing to zero-spin ground state (through intermediate state of spin one) were being established. The inequality of Bell was shown to be violated \cite{asp}. Since that time the Copenhagen quantum mechanics was being taken as the only theory of the microscopic world even if it contained some internal contradictions (denoted as quantum paradoxes), which maintains practically until now. It has been shown only very recently that the given conclusion has been based on several important mistakes as it will be explained in the following.

In \cref{sec:schr_eq_clas_phys} it will be shown that the Schr\"{o}dinger equation may be interpreted in full agreement with classical physics as far as it is not extended and non-classical characteristics are not added. It will be then shown in \cref{sec:mistaking_interpretation} that Bell inequality has been interpreted mistakenly as it does not hold in any probabilistic system, e.g., if randomly distributed spin orientations are measured (i.e., in the case of extended Schr\"{o}dinger equation). Some other aspects and problems will be mentioned in \cref{sec:problems}. Concluding remarks will be summarized in \cref{sec:conclusion}. 


\section{\label{sec:schr_eq_clas_phys}Schr\"{o}dinger equation and classical physics}

The Schr\"{o}dinger equation is defined by corresponding Hamiltonian
\begin{equation}
  i\hbar\frac{\partial}{\partial t}\psi(x,t)=H\psi(x,t), \;\;\;\;
     H=-\frac{\hbar^2}{2m}\triangle + V(x)   \label{schr}
\end{equation}
where time-independent Hamiltonian $H$ represents the total (kinetic and potential) energy of a given physical system and $x$ represents the coordinates of all corresponring matter objects. 
 Time evolution of physical quantities is expressed with the help of mean (or expected) values of corresponding operators: 
 \begin{equation}
           A(t) = \int\psi^*(x,t)\,A_{op}\,\psi(x,t)dx    \label{phy}
    \end{equation}
where functions $\psi(x,t)$ represent vectors and $A_{op}$ operators in some corresponding Hilbert space.

If a physical system is classical, i.e., if it may be described by Hamilton equations, then also any physical quantity $A(t)$ corresponds to a statistical combination of Hamiltonian solutions. The given problem has been discussed to greater details in earlier papers \cite{lok10,inte,scripta,intech,adv}. Here only main points will be summarized.  

Schr\"{o}dinger demonstrated in corresponding papers \cite{schr} that his equation gave the same results as the classical physics in the case of inertial motion; and it was assumed commonly that some different results would be obtained when non-zero potentials would be added. However, it has been shown by Hoyer \cite{hoyer} and Ioannidou \cite{ioan} that it has been possible to derive the Schr\"{o}dinger equation if together with the set of all basic solutions of Hamilton equations also the set of all statistical combinations of these solutions has been considered, limited only by a rather weak condition (e.g., Boltzmann statistics \cite{hoyer}). 

It has been then necessary to distinguish between the basic solutions of Eq.~(\ref{schr}) and their superpositions; the basic solutions (or states) being given by 
\begin{equation}
      \psi_E(x,t)=\psi_E(x)e^{-iEt}  
\end{equation}          
where $\psi_E(x)$ is eigenfunction of corresponding Hamiltonian 
\begin{equation}
      H \psi_E(x) \,=\,E\psi_E(x).  
\end{equation}
 The basic solutions of Schr\"{o}dinger equation (characterized always by one Hamiltonian eigenfunction only) represent the so called "pure" states and correspond to individual solutions of the Hamilton equations. 

Any solution of Schr\"{o}dinger equation corresponds then to a combination of basic solutions of Hamilton equations. However, it does not hold in opposite way: not each solution of Hamilton equations corresponds to a basic solution of  Schr\"{o}dinger equation as in the case of closed physical systems only the existence of some discrete set of states is allowed \cite{adv}. 

The Schr\"{o}dinger equation may be easily generalized to include also non-classical characteristics that have probabilistic character in individual physical events (spin orientations or similarly). The influence of this probabilistic behavior may be included into individual solutions of this equation while in the case of Hamilton equations very complicated combinations of individual solutions should have to be solved.   

   The Hilbert space must be then adapted to the characteristics of the corresponding physical system. The system of eigenstates is to be extended according to the set of additional characteristics. However, if the interpretation is to correspond to original proposal of Schr\"{o}dinger (respecting causal evolution of individual particles) the Hilbert space must consist of several subspaces (differing fundamentally from the Hilbert space required by Bohr). E.g., the Hilbert space representing the evolution of a two-particle system must be represented in the Hilbert space consisting at least of one pair of two mutually orthogonal subspaces  corresponding to incoming or outgoing states; see the mathematical analysis of Lax and Philips \cite{lax1,lax2} and also the last paragraph in \cref{sec:bell_limits}. The same Hilbert space structure followed if a particle pair came into being in the decay of an unstable particle and the decay law was asked to be purely exponential \cite{alda}. 


\section{\label{sec:mistaking_interpretation}Mistaking interpretation of Bell's inequality}

In the following we shall reproduce three different approaches, in which it will be shown that the inequality of Bell cannot hold in the corresponding probabilistic experiment. First, we shall discuss  two different approaches of Bell; then we shall show that the given probability combinations may posses three different limits according to different additional assumptions (defining different physical alternatives); and finally we shall call the attention to the fact that Bell's inequality represents unreasoned additional limiting condition to general limiting conditions derived by Boole  \cite{boole} for any probabilistic system already 150 years ago.

\subsection {\label{sec:bell_approach}Two Bell's approaches}

Bell derived his inequality for the first time in 1964 \cite{bell}. However, he himself had probably some doubts and tried to derive it in another way, too \cite{bell2}. Some other approaches were then applied to by other authors.
All approaches proposed at that time have been summarized and reproduced systematically by Clauser and Shimony \cite{clauser} in 1978.

The conclusion of the first approach presented originally in Ref.~\cite{bell} was based on the condition relating some probability values on different sides of the given experiment  (see Eqs.~(3.2) and (3.6) of Ref.~\cite{clauser}). The given condition required in principle that the transmission probabilities obtained on both sides of coincidence experiment were uniquely correlated, which did not correspond to the probabilistic characteristics of polarization measurement. 

In the other case presented in Ref.~\cite{bell2} the given approach was generalized and coincidence probabilities were defined as 
\begin{equation}
  P_{\alpha,\beta} = \oint \text{d}\lambda\, a_{\alpha}(\lambda)b_{\beta}(\lambda) 
\end{equation} 
where $\lambda$ represented corresponding spin orientations (randomly distributed). Bell's approach started practically (see Eqs.~(3.11-3.12) of Ref.~\cite{clauser}) from equation   
\begin{equation}
      P_{\alpha_1,\beta_1} - P_{\alpha_1,\beta_2} = \oint \text{d}\lambda \, [a_{\alpha_1}(\lambda)b_{\beta_1}(\lambda) - a_{\alpha_1}(\lambda)b_{\beta_2}(\lambda) ]. \label{Palfbet}
\end{equation} 
 Then the expression
\begin{equation}
 \oint \text{d}\lambda \, 
 [a_{\alpha_1}(\lambda)b_{\beta_1}(\lambda)  a_{\alpha_2}(\lambda)b_{\beta_2}(\lambda) - a_{\alpha_1}(\lambda)b_{\beta_2}(\lambda) a_{\alpha_2}(\lambda)b_{\beta_1}(\lambda) ] \label{zer} 
\end{equation} 
was added (or subtracted) to the right side of Eq.~(\ref{Palfbet}). The expression (\ref{zer}) equaled mathematically zero; however, it had important impact to final results as in the following calculations the changed order of individual probabilities influenced strongly corresponding coincidence probabilities (especially when some limit value for coincidence probability combination was established in the given approach). It was required in principle to hold $P_{\alpha,\beta_1}=P_{\alpha,\beta_2}$ for any corresponding triple of polarizer orientations. It means that coincidence probabilities were assumed to remain the same after interchanging polarizer orientations on one side while the orientation on the other side remained unchanged.  
Consequently, the polarization measurement was excluded and no place remained for the application of Bell's inequality.

It was argued that Bell's inequality was derived also with the help of other approaches (see \cite{clauser}). However, it was demonstrated  that some restraining condition has been always added (sometimes latently); see \cite{lok98} where the attempt to call the attention to the given inequality problem was undertaken for the first time. 

\subsection{\label{sec:bell_limits}Different limits of Bell's probability combination }

More detailed insight into the given problem may be obtained when one uses the approach proposed in Ref.~\cite{hill}. Individual quantities ($a_{\alpha},\,b_{\beta}$)  on the right side of Eq.~(\ref{bellprob}) are assumed now to represent operators corresponding to individual measurements; they act in the Hilbert space \footnote{It is very simplified Hilbert space representing measurement results only.}
   \begin{equation}
     {\mathcal H}\;=\; {\mathcal H}_a \otimes {\mathcal H}_b        \label{tens}
 \end{equation} 
where the subspaces ${\mathcal H}_a$ and ${\mathcal H}_b$  correspond to individual polarizers (detectors). It holds then
\begin{equation}
   0\;\leq\; |\langle a_{\alpha}\rangle| \leq \;1,  \;\;\;\;  0\;\leq\;|\langle b_{\beta}\rangle|\; \leq \;1\, .  
\end{equation}
The expectation values $|\langle B\rangle|$ of the Bell operator defined by Eq.~(\ref{bellprob})
may then possess  different upper limits according to the mutual commutation relations holding between the operators $a_j$ and $b_k$.
 
 It may be immediately seen that it must hold $\,<\!B^*B\!>\;\le 16\,$ and consequently $\,<\!B\!>\;\,\le\,4\,$. However, the limit 4  cannot be reached; in fact three lower limit may exist according to chosen commutation relations between operators $\,a_j\,$ and $\,b_j\,$, as shown already earlier (see \cite{lok05,lok09,nano}):
    \[ <\!B\!> \;\;\; \leq \;\;\;  2\sqrt{3}, \;\; 2\sqrt{2}\;\;\mathrm{or}\;\;2.  \]

\begin{itemize}
\item[({\it i})] The first limit $2\sqrt{3}$ (and actually the highest one, derived for the first time in \cite{revz}) corresponds to the case when no pair of the probability operators commutes
    $$  [a_j,b_k]\neq 0\;, \;\;\;[a_1,a_2]\neq 0, \;\;[b_1,b_2]\neq 0\, ,    $$
i.e., when the interaction at distance might exist as the operators from different Hilbert subspaces do not commute mutually.

\item[({\it ii})] The second limit $2\sqrt{2}$ corresponds to the bohmian (or of Schr\"{o}dinger) alternative, when only the operators belonging to the same subspaces commute (no interaction at distance or no entanglement exists); i.e., if
 $$ [a_j,b_k]= 0\;\;\;\;\mathrm {and}\;\;\;[a_1,a_2]\neq 0,\;[b_1,b_2]\neq 0\,.   $$

\item[({\it iii})] And finally, the third limit 2 corresponds to the classical case, when all operators $a_j$ and $b_k$ commute mutually, i.e., if
     $$ [a_j,b_k]= 0 \;,\;\;\;\;\;[a_1,a_2]\;=\;[b_1,b_2]\;= 0  $$
and the final coincidence results do not depend on spin orientations of detected objects, which corresponds to the original proposal of the Gedankenexperiment by Einstein.
\end{itemize}        

Only the classical alternative ({\it iii}) has been, therefore, excluded by the experimental results given in \cite{asp}. The other alternative ({\it ii}) corresponding to the experimental data represents then the case when some non-classical characteristics have been added and the given physical system is probabilistic; however, no entanglement existing. It is, of course, also the Copenhagen alternative ({\it i}) that has not been excluded on the basis of the given experimental results. 

However, there are serious arguments against the Copenhagen quantum mechanics; see summaries in \cite{lok09,inte}. Here we shall call the attention to one additional theoretical argument demonstrated on the example of three-dimensional harmonic oscillator; see \cite{osci}. It has been shown in the quoted paper that two important logical shortages presented by Pauli \cite{pauli} in 1933 and by Susskind and Glogower \cite {suss} in 1964  (attempts to solve the given problems having been done without any greater success in the end of the past century) could be removed when the physical quantities  of a closed system have been interpreted correctly; i.e., when the functions $\psi(x,t)$ and $\psi^*(x,t)$ (see Eqs.~(\ref{schr}) and (\ref{phy})) have been represented by vectors in the Hilbert space doubled two-times: firstly according to Lax and Phillips \cite{lax1,lax2} and also according to Fajn \cite{fajn}. Only in such a case the incoming and outgoing states will be distinguished and the evolution operator will be fully unitary as it is to be standardly required; the corresponding evolution being fully causal.


\subsection { Probabilistic physical systems}

 The probability problem existed, of course, in physics already in the preceding centuries when it was evident that an initial state could not be determined always quite exactly, e.g., the initial positions of individual matter objects. Often a very small position change could lead to significantly different evolution kinds; quite different results having been obtained with the help of Hamilton equations. Similar probabilistic extension followed then when some further internal characteristics (e.g., spin orientation) of individual matter objects have been added; as done by Bell having added spin measurements to Einstein's experiment. 

Already in 1854 Boole \cite{boole,vorob} derived for probabilities $p_1,p_2,....,p_n$ of some respective individual events $A_1,A_2,....,A_n$ the following inequalities for corresponding
probabilities $P(A_1\cup A_2\cup....\cup A_n)$ and $P(A_1\cap A_2\cap....\cap A_n)$
\begin{equation}
\max \{p_1,p_2,....,p_n\} \leq P(A_1\cup A_2\cup....\cup A_n)\leq   \min \{ 1, p_1+p_2+ ...... +p_n\} 
\label{eq:boole1}
\end{equation}
\begin{equation}
 \max \{0, p_1+p_2+....+p_n\; -n+1 \} \leq P(A_1\cap A_2\cap....\cap A_n)\leq  \min \{ 1, p_1, p_2, ......,p_n\} 
\label{eq:boole2} 
\end{equation}
According to Boole these inequalities also provide the best possible
estimates for probabilities $P(A_1\cup A_2\cup....\cup A_n)$ and
$P(A_1\cap A_2\cap....\cap A_n)$ if only the individual probabilities
$p_1,p_2,....,p_n$ are known.

Bell supposed that the probabilities for some different polarization orientations in the causal bohmian alternative might be mutually correlated. As introduced in \cref{sec:bell_approach}  he proposed two different assumptions leading to the same inequality.
 In 2004 Rosinger \cite{ros} showed then that a whole class of mathematical conditions might be chosen leading to the inequality derived by Bell for the given combination of coincidence probabilities; without any relation to a condition concerning physical nature. The given inequality should hold, of course, in addition to general inequalities derived by Boole. 
However, the corresponding probability correlations might be hardly acceptable in any real probability physical system.   

In the preceding a series of different arguments concerning the invalidity of Bell's inequality in corresponding coincidence experiments has been provided. Several remarks concerning the consequences will be mentioned in the following.

\section{\label{sec:problems} Some related problems and consequences}

It has followed from  preceding results that the Schr\"{o}dinger equation has represented realistic description of a physical system when the individual time-dependent solutions have been represented by vectors in the Hilbert space describing ordinary time-dependent evolution.
It means that if the classical evolution of physical quantities is to be reproduced the Hilbert space must consist of two or more mutually orthogonal subspaces (in contradiction to Bohr's assumption excluding such a possibility). 
However, even in such a case still one important problem more exists, concerning the attractive Coulomb potential between electron and proton. 

It has not been explained until now, e.g., why in the case of free proton and electron (being attracted by electric potential) hydrogen atom has been practically always formed if 
the mutual momentum of the given objects 
has been sufficiently low. It means that in addition to Coulomb interaction some short-ranged repulsive potential between the two given objects should exist. Such a situation may be phenomenologically described if the potential is suitably chosen. However, one may  hardly satisfactorily explain such a behavior with the help of point-like potential source if also the dimensions of hydrogen atom and proton are to be considered. 

The dimensions of proton and hydrogen atom according to contemporary theoretical and experimental results should be, however, so mutually different that the existing stability of hydrogen atoms in mutual low-energy collisions would be hardly acceptable, which opens the necessity to look for new kinds of explanation. It is also the idea of electrons running around the proton (or around atom nuclei) being attracted by some potential forces that may be hardly compatible with the stability of hydrogen atoms. It seems that some other kind of mutual forces should be taken into account. There  are already the short-ranged weak and strong forces that may be hardly described with the help of potential acting at distance; they 
should be probably interpreted as contact forces differing significantly from Coulomb forces acting evidently at distance (see \cite{intech}). 
A new description (and interpretation) of such contact forces should be looked for. 

In such a case, however, the dimensions of individual matter objects should play more substantial role. The dimensions being derived for protons on the basis of collision processes at very high energies at the present must seem to be too small if the protons should play the experimentally known role in corresponding matter structures.
 It seems to be necessary to assume that a strongly bound proton core is surrounded by some environment exhibiting much weaker interaction only; and to study whether this environment might be responsible for mutual interaction between individual fundamental matter objects at low energy values and for formation of atoms and also of crystals or solid substances. 

It is, of course, evident that the corresponding problems cannot be solved on the basis of mathematical formulas describing some phenomenological characteristics only. The description of physical systems must start from the ontological realistic basis of concerned objects. For the present, it will be necessary before all to study the dimensions of individual matter objects gained on the basis of collisions between different particles at divers energies (from high energies to very low values). The ontological collision model proposed recently in Prague \cite{intech} might be extremely advantageous as it contains only assumptions having clear physical meaning, which is not the case of standardly used phenomenological mathematical collision models.

  
\section{\label{sec:conclusion} Concluding remarks} 

Misleading interpretation of Bell's inequality has been surely the main source of deformed conclusions concerning the existence of entanglement between microscopic objects. However, the given physical picture in the recent past has been supported also by false conviction that the Schr\"{o}dinger equation led to the results differing significantly from classical physics. This equation has been, however, in fundamental agreement with classical ontological concept (which has been shown only recently); the misleading conviction has started from the fact that it has been possible to extend it easily so that it has involved and described non-classical characteristics, too. Consequently, the contemporary technological progress in microscopic region based also on Schr\"{o}dinger equation has not contradicted the classical ontological concept (requiring causality) as in the corresponding mathematical systems no assumptions added by Bohr have been applied to.
 
It might seem, therefore, that the Schr\"{o}dinger equation (leading to quantization of closed systems) might be applied in principle to all (microscopic and macroscopic) physical systems. However, it is not possible to explain, e.g., the emergence of stable bound states on the basis of standard forces acting at distance.
It seems to be necessary to consider the existence of some contact forces, too, acting directly between individual matter particles. In such a case it is necessary to take into account also the dimensions of individual objects that may be determined especially from elastic collision processes if a model enabling us to establish the dependences on collision impact parameters is used (see, e.g., \cite{intech}). 
  
However, if one takes the dimensions of proton determined in p-p collisions at very high energy values they differ fundamentally from the dimensions of atoms and their distances in solid substances (are much smaller). It might support the idea that  nucleon consists of a hard core held together by strong forces and is surrounded by greater sparser region exhibiting much weaker forces to other objects, which might open quite new ways in basic particle research.    
\\ [4mm]
 
{\footnotesize

}

\end{document}